\documentclass[showpacs,twocolumn,superscriptaddress,prl,preprintnumbers,nofootinbib]{revtex4}
\usepackage{amsmath,amssymb,bm,graphicx,color}

\begin{document}

\title{Heat Transport in Spin Chains with Weak Spin-Phonon Coupling}

\author{A. L. Chernyshev}
\affiliation{Department of Physics and Astronomy, University of California,
Irvine, California
92697, USA}
\author{A. V. Rozhkov}
\affiliation{Moscow Institute of Physics and Technology, Dolgoprudny,
Moscow Region, 141700 Russia}
\affiliation{Institute for Theoretical and Applied Electrodynamics, Russian
Academy of Sciences, Moscow, 125412 Russia}
\date{\today}
\begin{abstract}
The heat transport in a  system of $S\!=\!1/2$ large-$J$ Heisenberg  spin chains, 
describing  closely Sr$_2$CuO$_3$ and SrCuO$_2$ cuprates, is studied theoretically at $T\!\ll\! J$
by considering interactions of the bosonized spin excitations  with optical phonons and defects. 
Treating rigorously the multi-boson processes, 
we derive a microscopic spin-phonon scattering rate that adheres to an intuitive picture of phonons acting as 
thermally populated defects for the fast  spin excitations. The  mean-free path of the latter 
exhibits a distinctive $T$-dependence reflecting a critical nature of 
spin chains  and gives a close description of experiments. 
By the naturalness criterion of realistically small spin-phonon interaction, 
our approach stands out from previous considerations that  
require large coupling constants to explain the data and thus imply 
a spin-Peierls transition, absent in real materials.
\end{abstract}

\maketitle

The one-dimensional (1D) spin chains are among the first
strongly-interacting quantum many-body systems ever studied
\cite{JW,Bethe}
and they remain a fertile ground for  new
ideas~\cite{MBlocalization}
and for developments of advanced theoretical and numerical~\cite{boso, DMRG}
methods. A number of  physical realizations of spin-chain materials~\cite{Nagler,cuprates,Zheludev,Stone,Mourigal} 
have allowed for an unprecedentedly comprehensive 
comparisons between theory, numerical approaches, and experimental data
\cite{Giamarchi_book,JSCaux,AdrianF}.
Current theoretical challenges for these systems include their dynamical,
non-equilibrium, and transport properties~\cite{Fabian2009,Fabian2011,Sirker09,Sirker11,Cabra02,Brenig10,Fabian2015}. 
The transport phenomena are particularly challenging 
as the couplings to phonons and impurities, perturbations that are 
extrinsic to the often integrable spin systems, 
become crucial~\cite{ZNP97,Brenig_review,Zotos_review,Shimshoni03,Zotos06,RC}.
 
In this Letter, we address the problem of heat transport in 1D spin-chain
systems  by considering  coupling of spins to  optical phonons and
impurities and 
having in mind a systematic experimental thermal conductivity study in the  high-quality, 
single-crystalline,  large-$J$ spin-chain cuprates  Sr$_2$CuO$_3$ and SrCuO$_2$   that has been recently 
conducted \cite{Hess1,Hess2,Hess3,Hess4}. Several attempts to develop a suitable formalism 
to describe this phenomenon have been made in the past~\cite{RC,Shimshoni03,Zotos06}.
 However, these approaches either relied on unrealistic choices of parameters~\cite{RC,Shimshoni03}
or offered only qualitative insights~\cite{Shimshoni03,Zotos06}.

Below we attempt to bridge the gap between  experiment and theory.
We argue that the heat conductivity by spin excitations can be quantitatively described 
within the bosonization framework with the large-momentum scattering  
by optical phonons or impurities.
For  weak impurities, scattering grows stronger at lower  temperature, 
a feature intimately related to a critical character  of the $S\!=\!1/2$ Heisenberg chains  \cite{RC}. 
Taking into account multi-spin-boson processes, it follows naturally from our microscopic  calculations that 
 scattering  by phonons  bears a close similarity to that by weak impurities, 
 except that the phonons are thermally populated and thus control  heat transport at high $T$. 
 This is also in accord with a  physical picture of phonons playing the role of  impurities for the fast spin excitations. 
 Within this picture, the transport relaxation time  is  the same as 
spin-boson scattering time and the corresponding mean-free path fits excellently the available experimental data.  
Further systematic extensions of our theory to include multi-phonon scattering that can influence thermal 
conductivity at higher temperature are  briefly discussed.

Finally, we emphasize an important physical constraint on the strength of 
spin-phonon coupling of magnetoelastic nature~\cite{Bramwell90,CB}, which is weak 
in the materials of interest. While an estimate of this coupling can be made microscopically, 
a simple piece of phenomenological evidence for this criterion is the absence of the spin-Peierls transition 
in real compounds down to very low temperatures. 
Our theory easily satisfies the proposed constraint, 
setting itself apart from the previous approaches~\cite{RC,Shimshoni03}.
We thus provide  a  microscopic, internally consistent 
description of thermal transport and scattering in 1D spin chains, 
which satisfies naturalness criteria by having weak spin-phonon coupling 
and conforming to an analogy between  phonon and impurity scatterings.

\begin{figure}[tb]
\includegraphics[width=0.999\columnwidth]{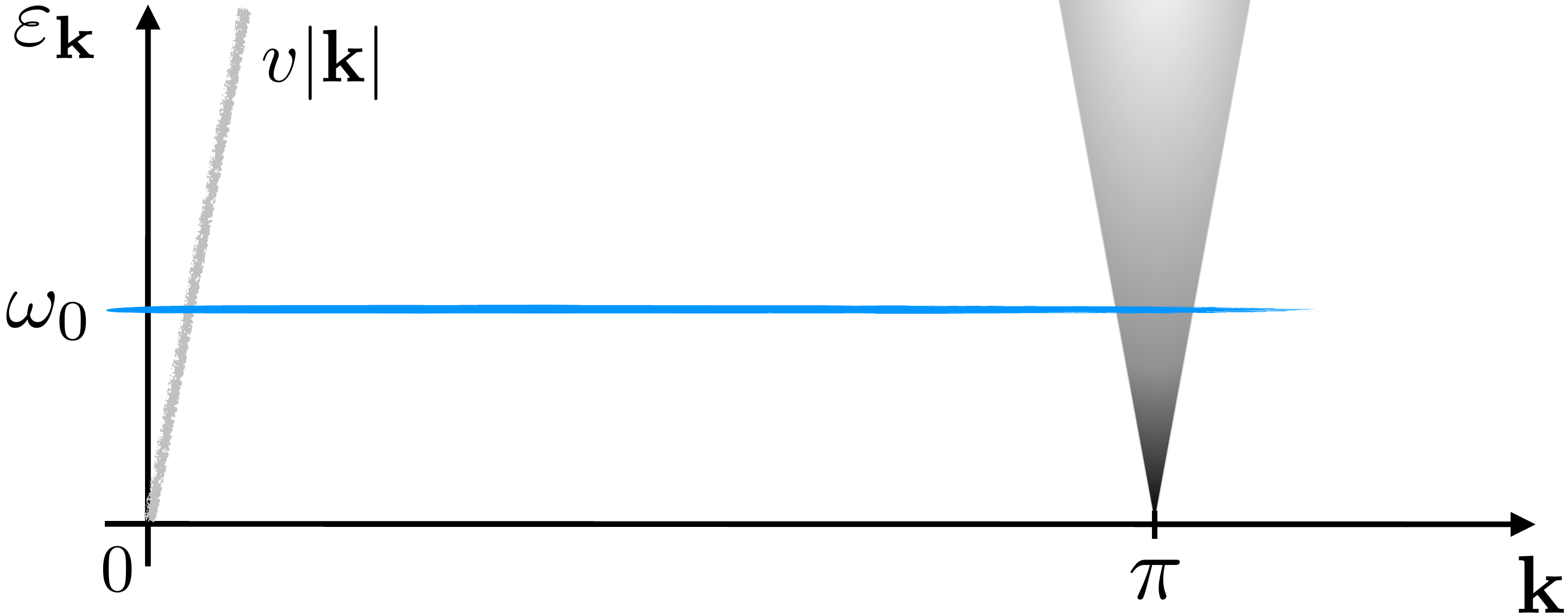}
\caption{(Color online)\ \  
Schematics of the spectra of bosonic excitations 
in a large-$J$, $S\!=\!1/2$ Heisenberg spin chain (dispersive branch
$\varepsilon_{\bf k}\!=\!v|{\bf k}|$ and a continuum at ${\bf Q}\!=\!\pi$) 
and of the dispersionless optical phonon branch $\omega_0$
(the horizontal line).  
\label{Fig:1}
}
\vskip -0.3cm
\end{figure}

\emph{Spin-phonon coupling Hamiltonian.}---%
The nearest-neighbor Hamiltonian of an $S\!=\!1/2$ Heisenberg chain magnetoelastically coupled to 
phonons  is
\begin{equation}
{\cal H}=\sum_{\langle ij\rangle} J\left({\bf r}_i-{\bf r}_j\right) {\bf S}_i\cdot{\bf S}_j  \, ,
\label{H} 
\end{equation}
\vskip -0.15cm \noindent
where
$\langle ij\rangle$
denotes nearest-neighbor lattice sites. A standard Jordan-Wigner
transformation with the subsequent bosonization \cite{Giamarchi_book}
and the lowest-order expansion in lattice displacements brings it to the
following form:
\begin{equation}
{\cal H}=\sum_{k}\varepsilon_{k}b^\dag_{k}b^{\phantom\dag}_{k} + {\cal H}_{\rm s-ph}\, ,
\label{Hk}
\end{equation}
where $b^{(\dag)}_{k}$ represents spin-boson operators of the excitation 
with  $\varepsilon_{k}\!=\!v|k|$  (sketched in Fig.~\ref{Fig:1}), velocity is $v\!=\!\pi Ja/2$,
$k$ is the 1D momentum, and $a$ is the lattice spacing. Hamiltonian 
${\cal H}_{\rm s-ph}$ describes a large-momentum,
$q\!\approx\!Q=\pi/a$, spin-boson scattering  by phonons
\begin{eqnarray}
{\cal H}_{\rm s-ph}=\frac{2\lambda}{\pi a^2} \int dx\, {\bf U}_x(x)\cos \Big(\hat \Phi(x)+Q x\Big)\, ,
\label{Hsp}
\end{eqnarray}
where $\lambda\!=\!a\partial J/\partial x$, $x$ is the direction along the chains, the lattice displacement field
${\bf U}(x)$ is associated with the optical and zone boundary phonons, and the spin-boson field 
$\hat \Phi(x)\!=\!\sqrt{\pi}\sum_k e^{ikx}(b^\dag_{k}+b^{\phantom\dag}_{-k})/\sqrt{L|k|}$, 
in which $L$ is the linear size of the chain and we used  the
Luttinger-liquid parameter ${\cal K}\!=\!1/2$ for the Heisenberg case \cite{supp}.
Small-momentum scattering is
deliberately ignored, as the corresponding vertex carries small in-plane
momentum of the phonon and leads to negligible scattering
effects~\cite{RC}. 

We note that boson-boson scattering cannot dissipate the heat current \cite{ZNP97,Brenig_review,Rosch06}
and thus is neglected.

\begin{figure}[t]
\includegraphics[width=0.999\columnwidth]{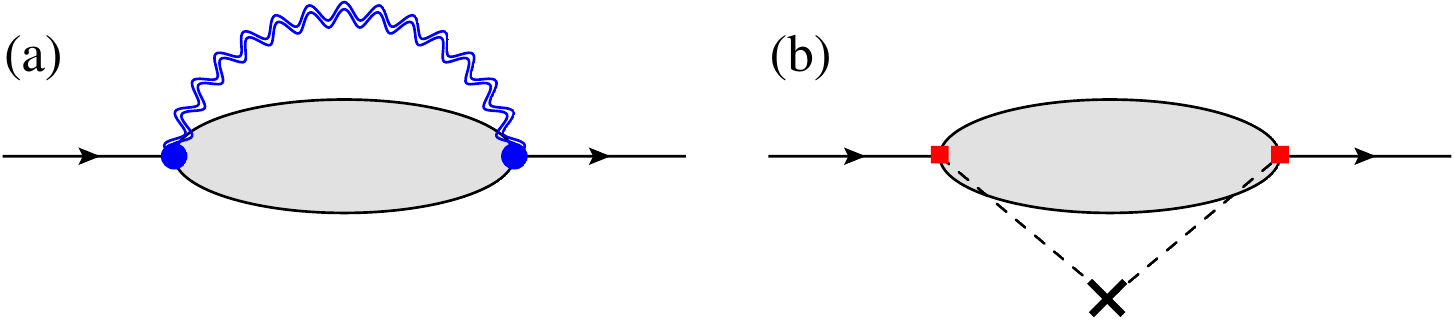}
\vskip -0.1cm
\caption{(Color online)\ \  
Multi-boson diagrams contributing to the scattering rate of spin bosons 
on (a) phonons and (b) weak impurities. 
Shaded ellipses represent a set of diagrams involving an arbitrary number of spin bosons in the intermediate state.
Solid and wavy lines are the Green's functions of spin bosons and phonons, respectively. 
}
\label{Fig:2}
\vskip -0.3cm
\end{figure}

\emph{Self-energy and relaxation rate.}---%
Assuming the spin-phonon coupling to be  small, a conjecture discussed below in detail \cite{CB}, 
one can consider only the second-order spin-boson  self-energy in Fig.~\ref{Fig:2}(a)  given by
\begin{eqnarray}
\Sigma_k (\tau)= - \frac{2\lambda^2}{\pi a^4 |k|}\int dx\, e^{ikx} 
D(\tau,x)
\big\langle  e^{ - i \hat \Phi(0,0) }e^{ i \hat \Phi(\tau,x) }\big\rangle\, ,
\label{Sigma1}
\end{eqnarray}
where $D(\tau,x)=\left\langle {\bf U}_x(0,0) {\bf U}_x(\tau,x) \right\rangle$ 
is the phonon propagator and the second quantization of lattice-displacement field is standard \cite{Ziman}. 
We exploit the large value  of $J$
compared to a typical Debye energy (in cuprates $J/\Theta_D\!\sim\!10$),
which allows us to neglect dispersion of the phonon branches near the $\pi$-point in Fig.~\ref{Fig:1}.
Then, the lattice-displacement correlator is fully local in space \cite{supp} and
separates into a sum over phonon branches $\ell$ that have non-zero projections of 
their polarizations, ${\bm \xi}^x_{\bf q\ell}$, on the chain axis $x$.
Considering for simplicity only one longitudinal phonon with the energy $\omega_0$ (see Fig.~\ref{Fig:1}), 
and reserving the right to 
add more phonon branches later, we obtain $D(\tau, x)
=a \delta (x)F_\tau(\omega_0)/2 m \omega_{0}$
with
\begin{eqnarray} 
F_\tau(\omega_0)=
n_0 e^{\omega_{0} \tau}+
(n_0 + 1) e^{-\omega_{0} \tau}
,\ \ \ \ 
\label{phonon}
\end{eqnarray}
where  $n_0\!=\!1/(e^{\omega_0/T}-1)$ is the phonon distribution function, $m$ is the mass of the unit cell, and 
$\hbar\!=\!k_B\!=\!1$.
 
For the bosonic field correlator in the spin-phonon self-energy (\ref{Sigma1}) in 
Fig.~\ref{Fig:2}(a), we note an immediate similarity to the second-order $T$-matrix for the weak impurity 
scattering in Fig.~\ref{Fig:2}(b), which also  generates a large-momentum
transfer \cite{RC}. 
The correlator  can be evaluated at $x\!\rightarrow\! 0$ and  $T\!\ll\!J$ \cite{supp,RC} and leads to 
\begin{eqnarray}
\big\langle e^{ - i \hat \Phi(0,0) }e^{ i \hat  \Phi(\tau,0) }
\big\rangle \approx\frac{ \pi T}{ J |\sin(\pi T \tau)| }.
\label{boson}
\end{eqnarray}
Then the self-energy at Matsubara frequency $\omega_n$ is
\begin{eqnarray}
\Sigma_{k}(\omega_n) = - g_{\rm sp}^2 \cdot \frac{2T J}{a |k|} 
\int_0^\beta d \tau\,
\frac{e^{i \omega_n \tau} - 1}{|\sin ( \pi T \tau )| }\, F_\tau(\omega_0),
\label{Sigma2}
\end{eqnarray} 
where we introduced a naturally appearing \emph{dimensionless} spin-phonon
coupling constant
$g_{\rm sp}\!=\!\lambda/(aJ\sqrt{2 m \omega_{0}})$~\cite{Bramwell90,CB}.
For the spin-boson scattering rate, we need  the imaginary part of the 
self-energy that is analytically continued to real frequencies.  The
transformations allowing us to perform the integration in Eq.~(\ref{Sigma2}) exactly
are delegated to Ref.~\cite{supp}. Here,  
we simply list the answer, 
\begin{eqnarray}
{\rm Im}\Sigma_{k}(\omega)
=
-g_{\rm sp}^2\, \frac{2J}{a |k|} 
\Big(2n_0+1\Big)\Big(1-f_+-f_-\Big)\, ,
\label{ImSigma}
\end{eqnarray} 
where  
$f_\pm\!=\!1/(e^{\omega\pm\omega_{0}}+1)$.
The fermionic distributions can be seen 
as a result of a re-fermionization of bosons via a 
multiple-boson scattering. 
The result~(\ref{ImSigma}) can be expanded in $\omega/T$, yielding
\begin{eqnarray}
{\rm Im}\Sigma_{k}(\omega) \approx - g_{\rm sp}^2\, \frac{2J\omega}{a |k|T} 
\cdot \frac{1}{\sinh\left(\omega_0/T\right)},
\label{ImSigma1}
\end{eqnarray} 
which holds exceptionally well for all $\omega\!\alt\!T$ of interest.  
Generally, the single-particle scattering rate~(\ref{ImSigma1}) 
should differ from the transport relaxation rate, but for the impurity-like
scattering the two become the same.

\emph{Mean-free path.}---%
Then, the on-shell approximation, 
$\omega\!=\!\varepsilon_k$, in Eq.~(\ref{ImSigma1}) yields the 
inverse spin-boson mean-free path, 
$1/\ell\!=\!1/v\tau$, due to spin-phonon scattering:  
\begin{eqnarray}
\left(\frac{\ell_{\rm sp}}{a}\right)^{-1}
=g_{\rm sp}^2\, \frac{2J}{T} 
\cdot \frac{1}{\sinh\left(\omega_0/T\right)}\, .
\label{ell_sp}
\end{eqnarray} 
This result is $k$-independent and thus can be compared directly to the transport 
mean-free path extracted from thermal conductivity  data
\cite{Hess1,Hess3}.
We note that the $1/T$ prefactor in Eq.~(\ref{ell_sp}) is strongly reminiscent of the
result for the  
scattering on weak impurities \cite{RC,musyu,AffleckOshikawa}:
$\left(\ell_{\rm imp}/a\right)^{-1}\!=\!n_{\rm imp}\left(\delta J/J\right)^2(J/T)$,
where $n_{\rm imp}$ is the concentration of such impurities and $\delta J$ is the  
strength of   impurity potential. Clearly, this scattering  gets stronger with 
lowering $T$, down to the Kane-Fisher scale, $T_{\rm KF}\!\propto\!\delta J^2/J$,
below which weak impurity becomes a strong scatterer, similar to a chain break  \cite{KF}. 
This behavior is a  consequence of a critical character  of spin chains \cite{RC,footnote}. 
Since phonons should be seen as weak impurities by the fast spin excitations, it is 
natural that the spin-phonon scattering yields the same $1/T$ prefactor 
in Eq.~(\ref{ell_sp}). 

While the other thermal factor in (\ref{ell_sp}), $1/\sinh(\omega_0/T)$, 
does not coincide with the phonon population $n_0$, both have the same
high- and low-$T$ asymptotes. For $T\!\ll\!\omega_0$, the
mean-free path  (\ref{ell_sp}) exhibits activated behavior, $\ell_{\rm sp}\!\sim\!e^{\omega_0/T}$,
similar to the findings of other
works~\cite{Shimshoni03,Zotos06}. 

In addition to the considered scattering mechanisms, 
the low-$T$ spin thermal conductivity in real 
materials is limited by strong  defects that act like chain breaks~\cite{Hess1,Hess2,Hess4}. 
The corresponding mean-free path is an average  length of a defect-free chain segment, 
$1/\ell_{\rm b}\!=\!n_{\rm b}$,
where $n_{\rm b}$ is the concentration of these defects.

\begin{figure}[t]
\includegraphics[width=0.999\columnwidth]{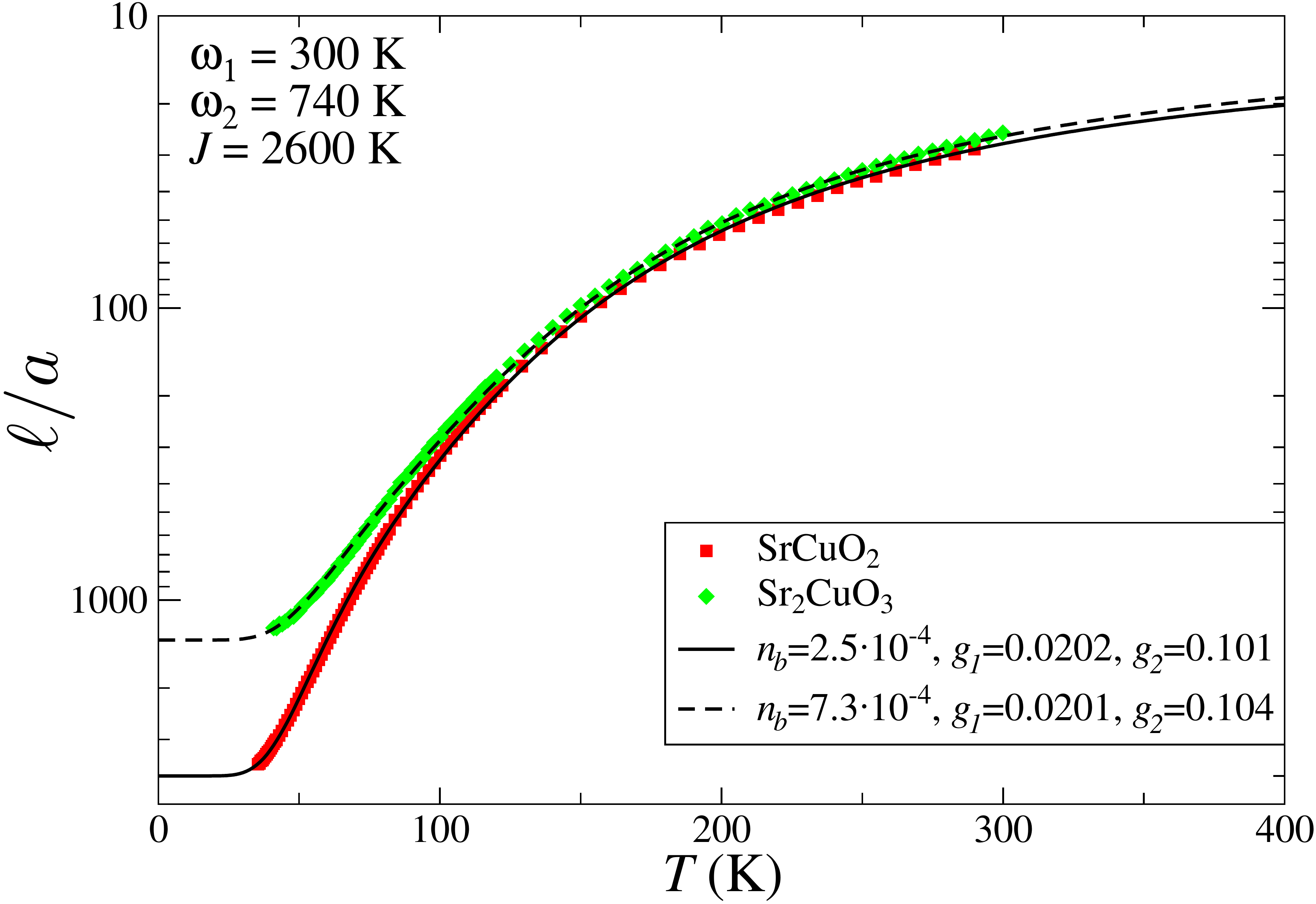}
\caption{(Color online)\ \  
Mean-free path of  spin excitations 
in Sr$_2$CuO$_3$ and SrCuO$_2$ \cite{Hess1,Hess2,Hess3} (the symbols). Lines are theory fits, see the text. 
Concentrations of strong impurities, $n_{\rm b}$, phonon energies, $\omega_{0,i}$, and spin-phonon coupling 
constants, $g_{i\rm,sp}$, are as indicated in the graph.
}
\label{Fig:3}
\vskip -0.3cm
\end{figure}

\emph{Comparison with experiments.}---%
Figure~\ref{Fig:3} shows the $T$-dependence of the mean-free path of  1D spin excitations 
in 
Sr$_2$CuO$_3$ and SrCuO$_2$ \cite{Hess1,Hess3}. 
The  data are extracted from the thermal conductivity measurements  
via a kinetic  relation, $\ell(T)\!=\!\kappa(T)/vC_V(T)$, using theoretical values  
 \cite{Klumper} for the specific heat of spin chains $C_V(T)$ [$\propto\!T$ at $T\!\ll\!J$].
Because of high purity, the mean-free path exceeds $10^3a$ at low $T$, with the difference between  
the two compounds due to the residual concentrations of the defects.
The two sets of data become quantitatively very close at higher $T$, 
implying that a similar scattering is dominating propagation of 
heat in both materials \cite{Hess3}.

Figure~\ref{Fig:3}  shows our successful fits of the data by combining spin-phonon (\ref{ell_sp}) 
and strong-impurity scatterings, 
$\ell^{-1}=\ell^{-1}_{\rm sp}+\ell^{-1}_{\rm b}$,  via Matthiessen's rule  \cite{Hess1}.
We note that the low-$T$ part of the data, $T\!\alt\!40$\,K, 
has a large uncertainty due to the subtraction of the phonon part of the thermal conductivity (see 
\cite{Hess1,Hess3}), and that it can be fit with an equal success by a 
combination of weak and strong impurities, $\ell^{-1}\!\approx\! \ell^{-1}_{\rm imp}\!+\! \ell^{-1}_{\rm b}$. 
Since it is a secondary issue for our study, the simplest 
account of this regime by strong impurities suffices. 
To fit the spin-boson mean-free path above 40\,K, we assume that the
spin-bosons are scattered  by two phonon modes with
$\omega_{0,1}\!=\!300$\,K and $\omega_{0,2}\!=\!740$\,K.
Of the two, the first roughly corresponds to the longitudinal
zone-boundary phonon and the second to the 
high-energy stretching mode~\cite{Pintschovius91,Ronnow14}, both likely having the strongest coupling to spin 
chains. In Eq.~(\ref{ell_sp}) we used the value of $J\!=\!2600$ K \cite{Sologubenko00} 
and the  spin-phonon coupling constants   $g_{\rm 1,sp}\!=\!0.020(1)$
and $g_{\rm 2,sp}\!=\!0.10(1)$ have provided the fit in Fig.~\ref{Fig:3} for Sr$_2$CuO$_3$ and SrCuO$_2$. 
By choosing different $\omega_{0,i}$'s, one can obtain somewhat different values of the $g_{\rm i,sp}$'s needed 
for the fit, but they never exceed or even reach the physical bounds discussed next.

\emph{Bounds on spin-phonon coupling.}---%
We now discuss physical bounds on the  spin-phonon coupling constant
$g_{\rm sp}\!=\!\lambda/(Ja\sqrt{2m\omega_0})$.
As discussed
in Refs.~\cite{Bramwell90,CB},
the constant is a product of two  parameters, with one characterizing the change
of $J$ by atomic displacement,
$\gamma\!=\!\lambda/J\!=\!a(\partial J/\partial x)/J$,
and the other representing an amplitude of  zero-point atomic motion relative to a
lattice constant~\cite{Ziman},
$\alpha\!=\!\hbar/\sqrt{2ma^2\omega_0}$, where $\hbar$ is made explicit and  
$m$ is the reduced mass associated with the phonon mode $\omega_0$. 
Parameter $\alpha$ is small, while $\gamma$ can be
large~\cite{Bramwell90,Ronnow14} 
because the superexchange is sensitive to
the interatomic distance. Regarding cuprates, one can estimate 
$\alpha\!\approx\!0.01$.
The superexchange parameter has a larger uncertainty, with indirect studies
giving a range of
$\gamma\!=\! 3\!-\!14$ \cite{Ronnow14,Takigawa} 
and a consideration of a wider class of materials suggesting an upper limit of $\gamma\!\leq\!20$ 
\cite{Bramwell90}.
Thus, the microscopic upper bound on the spin-phonon coupling constant in
1D cuprates can be put at
$g_{\rm sp}^{\rm max}\!\approx\!0.2$, 
justifying the weak-coupling treatment of the spin-boson scattering on
phonons
in Eq.~(\ref{Sigma1}).

A less restrictive, but   purely phenomenological criterion limiting the strength
of the spin-phonon coupling is the absence of the spin-Peierls transition in 1D cuprates  
down to about 5 K ($\approx\!0.002J$), where the 
3D  N\'eel ordering can be argued to preempt other transitions.
Using $T_{\rm P}\!\approx\! Je^{-1/g_{\rm sp}^2}$, this can be translated to the upper limit 
on the spin-phonon coupling  $g_{\rm sp}^{\rm max}\!\approx\!0.35$.

We now offer a critique of the previous   considerations of  thermal transport in 1D spin
chains. In particular, in experimental
works~\cite{Hess1,Hess3,Sologubenko00},
the spin-phonon mean-free path is repeatedly fit by the form $\ell_{\rm sp}^{-1}\!=\!ATe^{-\omega^*/T}$,
 with $\omega^*\!\approx\! 200$ K, inspired by the   phonon-mediated Umklapp scenario \cite{Shimshoni03}.
First, most of the data in Fig.~\ref{Fig:3} should be outside the quantitative accuracy range 
of this expression, which is limited to $T\!\alt\!\omega^*/3\!\approx\!70$\,K, 
as the exponent is only a low-$T$ limit of the phonon distribution function.   
More importantly, translating the values of $A$ used in Refs.~\cite{Hess1,Hess3,Sologubenko00} to the dimensionless 
coupling constant via $A\!=\!g_{\rm sp}^2/Ja$ gives $g_{\rm sp}\approx 1$, 
which is exceedingly large for the perturbative treatment to hold and lies
way outside the allowed range. This strong coupling also
implies a spin-Peierls transition at $T_{\rm P}\!\sim\! J$, while no such transition is 
observed. Likewise, our previous study, which considered small-momentum scattering
on acoustic phonon branches~\cite{RC}, required an anomalously 
strong spin-phonon interaction $g_{\rm sp}\!>\!1$. 
Thus, there is a serious ``naturalness'' problem with previous theoretical 
considerations. 

In the present work, the dimensionless spin-phonon coupling constants
are well within the range of the microscopic expectations for the cuprates, $g_{\rm sp}\!=\!0.02\!-\!0.1$, 
implying extremely low spin-Peierls transition temperatures. 
While the offered analysis of the physical bounds 
is not a proof of the validity of our theory, it is certainly
a strong argument against the validity of the previous approaches, 
which require unphysically large spin-phonon coupling.

\emph{Multi-phonon scattering.}---%
We note that for $T\!\agt\!\omega_0$ the spin-boson mean-free path in~(\ref{ell_sp}) saturates 
at $\left(\ell_{\rm sp}/a\right)^{-1}\!\approx\!  
g_{\rm sp}^2 2J/\omega_0$.
While this is not unphysical, one can still expect that the other, $T$-dependent terms may become important 
for $T\!\agt\!\omega_0$.
Corrections of the order $T/J$ are neglected in our derivation (see~\cite{supp}),
since $T/J$ is small in the relevant temperature range. 
Another possible source of the $T$-dependence is the multi-phonon scattering.
Superficially, the two-phonon scattering processes have to be negligible 
because of the smallness of the spin-phonon coupling discussed above.
However, there are factors that can compensate for this smallness. First, the two-phonon 
scattering is less restrictive, as the transverse phonons can also
contribute. Second, in the non-Bravais lattices, the  two-phonon processes
are also amplified by the number of atoms in a unit cell, $N_a$. That is,
for the single-phonon processes, the number of  longitudinal phonons that 
couple to spins via (\ref{Hsp}) is $N_a$, of which we have chosen only two
for our fits in
Fig.~\ref{Fig:3}.
On the other hand, when a spin-boson scattering is due to the 
emission or absorption of two phonons, the number of possible processes can
be as large as ${\cal O}(N_a^2)$.
A na\"{i}ve and certainly overly optimistic estimate of their number assuming
independent polarization and a branch index for each  phonon involved in the scattering 
yields $(3N_a)^2$.
In cuprates \cite{Pintschovius91}, the total number 
of phonon modes is large, so   this
combinatorial factor can be substantial.

A somewhat tedious, but straightforward algebra \cite{supp} yields the
following result for the two-phonon scattering
\begin{eqnarray}
\left(\frac{\ell_{\rm sp,2}}{a}\right)^{-1}
=g_{\rm sp,2}^2\, \frac{J}{T} 
\cdot \frac{\cosh\left(\omega_0/T\right)}{\sinh^2\left(\omega_0/T\right)},
\label{ell_sp2}
\end{eqnarray}
where $g_{\rm sp,2}^2\!\propto\!C_2g_{\rm sp}^4$. When compared to~(\ref{ell_sp}), 
the result in~(\ref{ell_sp2}) contains an extra factor, $g_{\rm sp}^2\!\sim\!0.01$,
and a large combinatorial factor, $C_2$. 
Clearly, at $T\!\ll\!\omega_0$, the two-phonon mean-free path follows the same behavior as (\ref{ell_sp}),
thus simply renormalizing single-phonon scattering. However, at $T\!\agt\!\omega_0$, it carries an extra 
power of $T/\omega_0$, $\left(\ell_{\rm sp,2}/a\right)^{-1}\!\approx\!g_{\rm sp, 2}^2 JT/\omega^2_0$, thus
amounting to an expansion in $T/\omega_{0,i}$, which can be argued to follow
naturally from the multi-phonon scattering processes.

\begin{figure}[t]
\includegraphics[width=0.999\columnwidth]{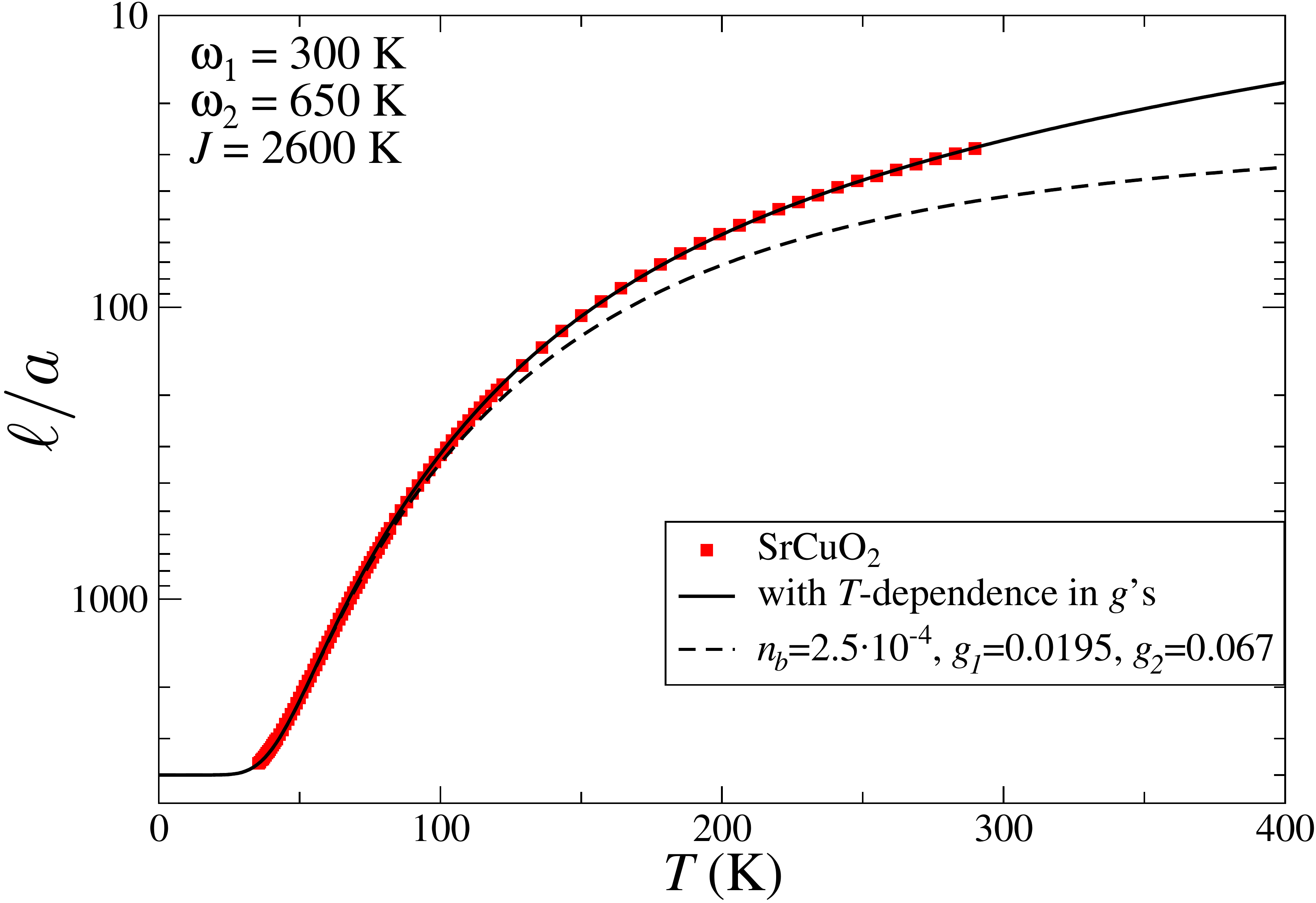}
\caption{(Color online)  
Same as in Fig.~\ref{Fig:3}. The solid line includes $T$-dependence in the spin-phonon coupling; see the text.
}
\label{Fig4}
\vskip -0.3cm
\end{figure}

Without going into non-generic microscopic considerations, one can suggest a simple ansatz to account for 
the $T/\omega_{0,i}$-expansion with the $T$-dependence of the spin-phonon coupling in the form 
$g_{\rm sp,i}(T)\!=\!g_{\rm sp,i}\left(1\!+\!r_i n_{0,i}\right)$, where $n_{0,i}\!=\!1/(e^{\omega_{0,i}/T}\!-\!1)$ 
as before. This form meets  both the low-$T$ and the high-$T$ behavior of the two-phonon 
mean-free path discussed above. A fit of the SrCuO$_2$ data using this ansatz with 
$r_i\!=\!1$ is provided in Fig.~\ref{Fig4}. The bare 
spin-phonon coupling constants $g_{i,\rm sp}$ are even smaller than in Fig.~\ref{Fig:3}, especially for the 
higher-energy mode. The result with the bare $g_{i,\rm sp}$'s 
is provided for comparison. Although this figure is an illustration 
showing that our theory allows for systematic extensions 
by including multi-phonon processes, 
it also demonstrates a  potential role of the latter in the $T\!\agt\!\omega_{0}$ regime 
and thus contributes to the general description of the heat transport in 
spin-chain materials.

\emph{Conclusions.}---%
We  have provided a   consistent microscopic theory for thermal transport and scattering in 
1D spin chains  which   stands out from previous attempts at such a theory   
by having weak spin-phonon coupling and conforming to the analogy of the phonon scattering to that on 
impurities. 
We have successfully fit the available experimental data and discussed possible extensions of our theory
for higher-$T$. Our approach should be  applicable to the thermal conductivity 
in spin-ladder materials and can be extended to the transport phenomena in  
a variety of Luttinger liquids and ultracold atomic gases. Numerical verification of our results is also called for.

\begin{acknowledgments}
\emph{Acknowledgments.}---%
We would like to thank Christian Hess for sharing his previously published
data and for the numerous fruitful and enlightening discussions.  A.~L.~C. also
thanks Andrey Zheludev for his patience, unabating constructive criticism,
and useful suggestions.  Work of A.~L.~C. was supported by the U.S.
Department of Energy, Office of
Science, Basic Energy Sciences under Award No. DE-FG02-04ER46174.
A.~L.~C. would like to thank KITP UC Santa Barbara, where part of this
work was done, for their hospitality. The work at KITP was supported in part by
 the National Science Foundation under Grant No. NSF PHY11-25915.
A.~V.~R. acknowledges the support of RFBR through Grants No. 14-02-00276-a and  No. 15-02-02128, and
of the Russian Science Support Foundation.
\end{acknowledgments}


\newpage \
\newpage
\onecolumngrid
\begin{center}
\vskip -0.8cm
{\large\bf Heat Transport in Spin Chains with Weak Spin-Phonon Coupling: \\ 
Supplemental Material}\\ 
\vskip0.35cm
A. L. Chernyshev$^1$ and A. V. Rozhkov$^{2,3}$\\
\vskip0.15cm
{\it \small $^1$Department of Physics and Astronomy, University of California, Irvine, California
92697, USA\\
$^2$Moscow Institute of Physics and Technology, Dolgoprudny,
Moscow Region, 141700 Russia\\
$^3$Institute for Theoretical and Applied Electrodynamics, \\ Russian
Academy of Sciences, Moscow, 125412 Russia}\\
{\small (Dated: December 31, 2015)}\\
\vskip 0.1cm \
\end{center}
\twocolumngrid
\renewcommand{\theequation}{S\arabic{equation}}
\setcounter{equation}{0}

\section{Spin-boson scattering on phonons}

\subsection{Bosonization and coupling to phonons}

Bosonization of the Jordan-Wigner fermion operators for the $S=1/2$ chain 
proceeds via introduction of the bosonic fields $\Phi$ and $\Theta$ 
for the right- and left-moving fermion operators \cite{bosos}
according to
\begin{eqnarray}
\psi_{\rm R} 
=
\frac{1}{\sqrt{2\pi a}} 
e^{i\sqrt{\pi} ( \Phi + \Theta )},
\\
\psi_{\rm L} 
=
\frac{1}{\sqrt{2\pi a}} 
e^{i\sqrt{\pi} ( - \Phi + \Theta )}.
\end{eqnarray}
The bilinear combination of the fermionic operators
$\psi^\dag_{\rm R} \psi^{\vphantom{\dagger}}_{\rm L}$
can be expressed as follows
\begin{eqnarray} 
\psi^\dag_{\rm R} \psi^{\vphantom{\dagger}}_{\rm L}
=
\frac{1}{2\pi a}
e^{ - i \sqrt{4\pi} \Phi }
=
\frac{1}{2\pi a}
e^{ - i \sqrt{2\pi} \tilde \Phi },
\end{eqnarray}
where the rescaled bosonic field
$\tilde \Phi = (1/\sqrt{\cal K}) \Phi = \sqrt{2} \Phi$.
The Tomonaga-Luttinger parameter
${\cal K} = 1/2$
corresponds to the Heisenberg limit of the model.

Coupling to phonon field that leads to the large-momentum scattering can be written as
\begin{eqnarray}
\label{H_U}
{\cal H}_{\rm s-ph}
=
\frac{\lambda}{\pi a^2} \int dx {\bf U}_x(x)
e^{ \sqrt{2\pi} \tilde \Phi(x) - i \pi x/a}
+
\text{h.c.},
\end{eqnarray}
where
$\lambda\!=\!a\partial J/\partial x$
and
${\bf U}_x (x)$ 
is the $x$-component of the lattice displacement field
${\bf U}$
for the optical and zone boundary phonons.  

\subsection{Spin-boson self-energy}

Since we are interested in the scattering involving momenta close to the 1D zone boundary momentum
$Q=\pi/a$, one can assume 
that the corresponding phonon has zero velocity. Then we can write the spin-boson self-energy as
\begin{eqnarray}
\Sigma_k (\tau)
=
- \frac{2\lambda^2}{\pi a^4 |k|}
\int dx e^{ikx}
\left\langle {\bf U}_x(0,0) {\bf U}_x (\tau,x) \right\rangle
\label{Sigma_s}
\\
\nonumber 
\times\left\langle 
e^{ - i \sqrt{2\pi} \tilde \Phi(0,0) }
e^{ i \sqrt{2\pi} \tilde \Phi(\tau,x) }
\right\rangle.
\end{eqnarray}

\subsubsection{Phonon correlation function}

Using  standard quantization of the displacement field,  phonon correlation function in (\ref{Sigma_s}) is
\begin{eqnarray}
\label{propag3D}
&&\left\langle {\bf U}_x(0,0) {\bf U}_x (\tau, {\bf R}) \right\rangle
=
\frac{1}{N}
\sum_{{\bf P} \ell}
	\frac{e^{-i{\bf P} {\bf R} }}
             {2 m \omega_{0 \ell}}
	(\xi^x_{{\bf P} \ell})^2
\\
\nonumber 
&&\phantom{\langle {\bf U}_x(0,0)}
\times	\left(
		\left\langle a^\dag_{{\bf P} \ell} (0)
			a^{\vphantom{\dagger}}_{{\bf P} \ell}(\tau)
		\right\rangle
		+
		\left\langle a^{\vphantom{\dagger}}_{-{\bf P} \ell}(0)
			a^\dag_{-{\bf P} \ell} (\tau)
		\right\rangle
	\right).
\end{eqnarray}
Assuming, for simplicity, that the dynamical force matrix at a given
momenta can be diagonalized so that one of the polarizations (longitudinal)
is along the chain and two remaining ones are 
perpendicular, then all of the relevant polarizations yield
$(\xi^x_{{\bf P} \ell})^2=1$
and the summation in (\ref{propag3D}) will be over $N_a$ independent
longitudinal phonon modes with individual frequencies $\omega_{0,\ell}$,
where $N_a$ is the number of atoms per unit cell.

If one were to include dispersion of  phonons into consideration, the phonon propagator  in (\ref{propag3D}) 
will not be strictly local. However, for the fast spin excitations, the phonon does not propagate
far during the time of interaction.
Thus, the effect of the phonon dispersion is small by the small parameter $\Theta_D/J\sim 0.1$ and our 
$\omega_{0,\ell}$ correspond to 
the average energies of the phonon branches that couple strongly to spins.

Consider coupling to an individual phonon mode with the energy $\omega_0$.
For this case we derive
\begin{eqnarray} 
\left\langle {\bf U}_x(0,0) {\bf U}_x (\tau, {\bf R}) \right\rangle
&=&
\frac{V_0 \delta ({\bf R})}
             {2 m \omega_{0}}
\\
\nonumber
&\times&
	\left[
		n_0 e^{\omega_{0} \tau}
		+
		(n_0 + 1) e^{-\omega_{0} \tau}
	\right],
\\
\mbox{with}\quad n_0&=&\frac{1}{e^{\omega_0/T} - 1},
\label{eq::n0_def}
\end{eqnarray} 
where $V_0$
is the volume of the unit cell, 
$n_0$
is the phonon occupation number, and we assumed that 
$\tau\! >\! 0$.
Since we need to evaluate the phonon propagator on a single chain, one can
write
$V_0 \delta({\bf R})\!
= \!
a \delta(x)$,
which  leads to
\begin{eqnarray} 
\label{ph_propag}
&&\left\langle {\bf U}_x(0,0) {\bf U}_x (\tau, x) \right\rangle
\\
\nonumber 
&&\phantom{\langle {\bf U}_x(0,0)}
=	\frac{a \delta (x)}
             {2 m \omega_{0}}
	\left[
		n_0 e^{\omega_{0} \tau}
		+
		(n_0 + 1) e^{-\omega_{0} \tau}
	\right].
\end{eqnarray}
Obviously, for the case of $N_a$ phonon modes, one  needs to reinstate the
summation over them.

\subsubsection{Spin-boson correlation function}

We now turn to the spin-boson part. We have previously obtained the spin-boson correlation function in \cite{CRs}. 
Here we provide a more rigorous derivation of it. First,
\begin{eqnarray}
\left\langle 
e^{ - i \sqrt{2\pi} \tilde \Phi(0,0) }
e^{ i \sqrt{2\pi} \tilde \Phi(\tau,0) }
\right\rangle
=
e^{
		-\pi \left\langle 
			\left[
				\tilde \Phi (0, 0)
				-
				\tilde \Phi (\tau, 0)
			\right]^2
			\right\rangle
}.\quad
\end{eqnarray}
Thus, we are to calculate 
$g(\tau)$
defined by
\begin{eqnarray}
&&g(\tau)
=
-\left\langle 
		\left[
			\tilde \Phi (0, 0)
			-
			\tilde \Phi (\tau, 0)
		\right]^2
\right\rangle
\\
\nonumber 
&&\phantom{g(\tau)}=
2
\left\langle 
			\tilde \Phi (0, 0)
			\tilde \Phi (\tau, 0)
\right\rangle
-
2
\left\langle 
			\tilde \Phi (0, 0)
			\tilde \Phi (0, 0)
\right\rangle.
\end{eqnarray}
Therefore, we have
\begin{eqnarray}
\label{summation}
g(\tau)
=
\frac{2T}{L}
\sum_{\omega_n }
	\left( e^{i \omega_n \tau } - 1 \right)
\sum_{k\ne 0}
	\frac{v}{\omega_n^2 + v^2 k^2},
\\
\nonumber
\mbox{with} \quad
\omega_n = 2 \pi T n,\ \ n\text{  = integer}.
\end{eqnarray}
Discarding the vanishing 
$\omega_n\!=\!0$
term in (\ref{summation}), 
we  have the following identity for any $\omega_n\!\neq\!0$ 
\begin{eqnarray}
&&g(\tau) 
=
\frac{2T}{L}
\sum_{\omega_n \ne 0}
	\left( e^{i \omega_n \tau } - 1 \right)
\sum_{k\ne 0}
	\frac{v}{\omega_n^2 + v^2 k^2}
\\
\nonumber 
&&\phantom{g(\tau)}
=2T
\sum_{\omega_n \ne 0}
\int_{-\infty}^{+\infty}
\frac{v dk}{2\pi}
	\frac{
		\left( e^{i \omega_n \tau } - 1 \right)
	     }
	     {\omega_n^2 + v^2 k^2}
	e^{-\pi a|k|/2}.
\end{eqnarray}
Here the exponent
$e^{-\pi a |k|/2}$
is used to set an ultraviolet cutoff. Next, we rewrite $g(\tau)$ as
\begin{eqnarray} 
g(\tau)
=
\frac{T}{\pi}
\int_{-\infty}^{+\infty}
	\frac{dy}{1 + y^2}
\sum_{\omega_n \ne 0}
	\frac{
		e^{i \omega_n \tau } - 1
	     }
	     {
		|\omega_n|
	     }
	e^{-y|\omega_n|/J},
\qquad
\end{eqnarray}
where $y$ is the dimensionless integration variable and we used
$v = \pi Ja/2$.
The summation over 
$\omega_n$
can be performed with the help of an identity
\begin{eqnarray}
\sum_{\omega_n > 0}
	\frac{
		e^{ \omega_n (i \tau - y/J)}
	     }
	     {
		\omega_n
	     }
=
-\frac{1}{2\pi T}
\ln
\left(
	1 - e^{2\pi T (i\tau - y/J)}
\right).
\quad
\end{eqnarray}
As a result we obtain
\begin{eqnarray}
&&g(\tau)=\frac{1}{2\pi^2} \int \frac{dy}{1+y^2}\nonumber\\
&&\phantom{g(\tau)=}
\times\ln \left[\frac{\left(1-e^{-2\pi Ty/J} \right)^2}{1+e^{-4\pi Ty/J}-2 e^{-2\pi Ty/J} \cos (2\pi T\tau )}\right]\nonumber\\
&&\phantom{g(\tau)}
\approx\frac{1}{2\pi^2}\int \frac{dy}{1+y^2}\ln \left[\frac{s y^2}{s y^2 + 4 \sin^2(\pi T \tau)}\right],\quad\quad\quad
\end{eqnarray} 
where
$s = (2 \pi T/J)^2 \ll 1$.
The remaining integral can be found in \cite{GR1}, which gives
\begin{eqnarray}
g(\tau)
=
\frac{1}{\pi}
\ln \frac{\sqrt{s}}{2|\sin(\pi T \tau)| + \sqrt{s}}\, .
\end{eqnarray}
Exponentiating $g(\tau)$, we arrive to the following expression for the spin-boson correlator in (\ref{Sigma_s})
\begin{eqnarray}
\label{eq::spin_corr}
&&\left\langle 
e^{ - i \sqrt{2\pi} \tilde \Phi(0,0) }
e^{ i \sqrt{2\pi} \tilde \Phi(\tau,0) }
\right\rangle
\\
\nonumber
&&\phantom{\langle e^{ - i \sqrt{2\pi}}}
=\frac{2 \pi T}{2 J |\sin(\pi T \tau)| + 2 \pi T}
\approx
\frac{ \pi T}{ J |\sin(\pi T \tau)| }\, ,
\end{eqnarray}
where in the last expression we neglected contributions of order
${\cal O}(T/J)$.

\subsubsection{Evaluation of the self-energy}

Using the phonon propagator (\ref{ph_propag}) and the spin-boson correlation function (\ref{eq::spin_corr})
obtained above and transforming the  spin-boson self-energy
$\Sigma_k(\tau)$ in (\ref{Sigma_s}) to the Matsubara 
frequency domain yields $\Sigma_k(\omega_n)$ given by
\begin{eqnarray}
\label{Sigma_om}
&&\Sigma_{k}(\omega_n)
= - \left(
	\frac{\lambda^2}{2 m a^2\omega_{0}}
     \right)
\left( \frac{2T}{Ja |k|}\right)
\\
\nonumber
&&\times
\int_0^\beta d \tau
\frac{e^{i \omega_n \tau} - 1}{|\sin ( \pi T \tau )| }
	\left[
		n_0 e^{\omega_{0} \tau}
		+
		(n_0 + 1) e^{-\omega_{0} \tau}
	\right].
\end{eqnarray} 
Therefore, we need to evaluate the following integral
\begin{eqnarray}
\label{eq::def_integral}
I_{\varepsilon}(\omega_n)
=
\int_0^\beta d \tau
\frac{e^{i \omega_n \tau} - 1}{\sin ( \pi T \tau )}
		e^{\varepsilon \tau},
\end{eqnarray}
for 
$\varepsilon = \pm \omega_0$. Here we used that $\tau>0$.

\begin{figure}[b]
\includegraphics[width=0.5\columnwidth]{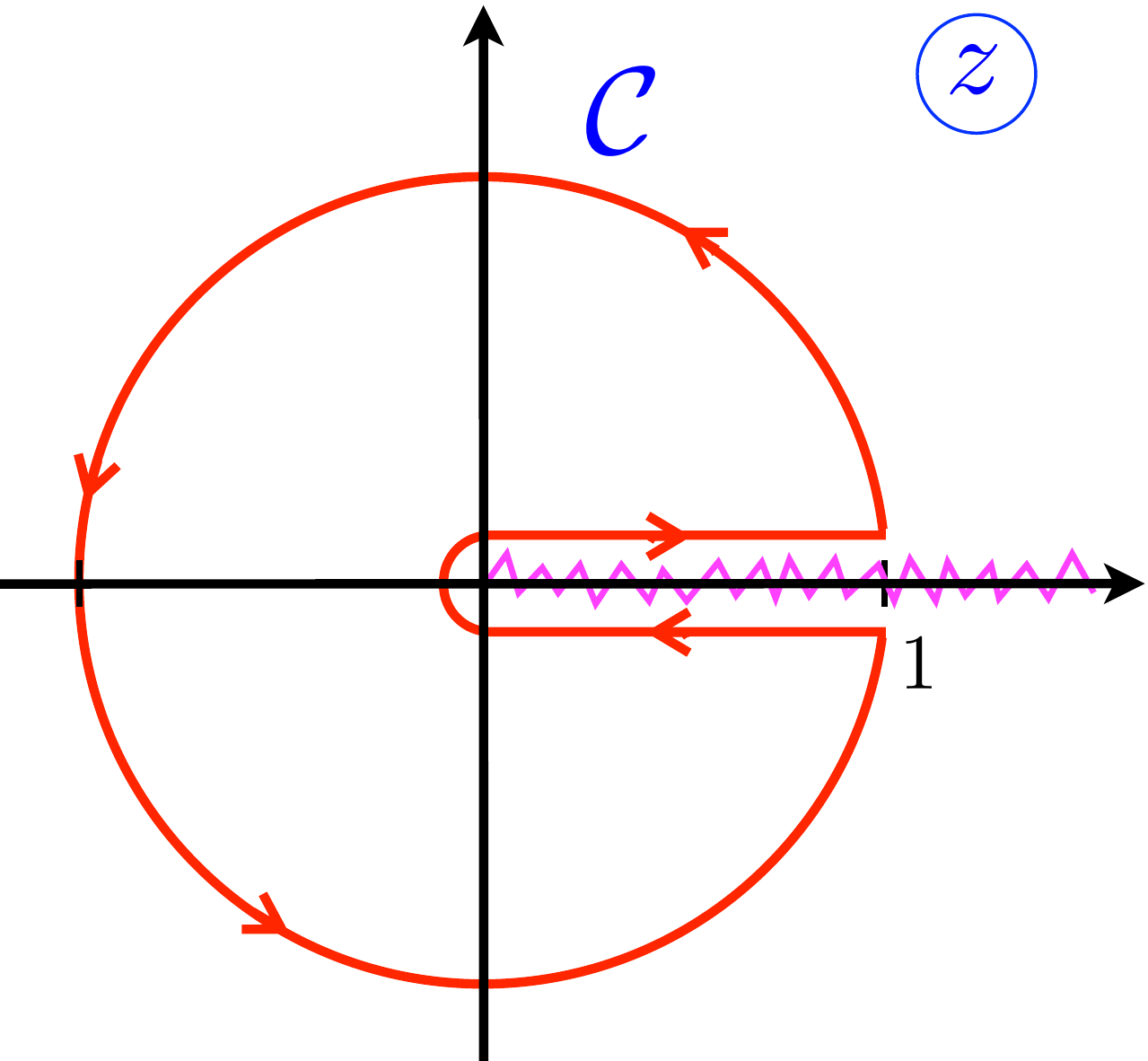}
\caption{(Color online)\ \  
Contour ${\cal C}$ for the evaluation of $I_{\omega_0}$.
}
\label{Fig:contour}
\vskip -0.3cm
\end{figure}

Consider
Eq.~(\ref{eq::def_integral}) 
for
$\varepsilon = \omega_0$.
Defining the complex variable 
$z = e^{i2\pi T \tau}$,
we transform 
$I_{\omega_0}$
into an integral along the unit circle in the $z$-plane. This contour,
however, is not closed, since the circle is cut at
$z = 1$ 
by a branch-cut running along the real axis from zero to 
$+\infty$.
Completing the unit circle to the contour
${\cal C}$,
see
Fig.~\ref{Fig:contour},
one can verify that
\begin{eqnarray}
I_{\omega_0}+I_{\omega_0}^r
=
\frac{1}{\pi T}
\int_{\cal C} dz \,\frac{1-z^n}{1-z} \, z^{-i \varphi_0 - 1/2}=0\,,
\end{eqnarray} 
where $\varphi_0 =\omega_0/2 \pi T$,
and
$I_{\omega_0}^r$
contains two segments of the contour
${\cal C}$
along the real axis.

Thus, $I_{\omega_0}$ reduces to the integral  of the real variable
\begin{eqnarray}
\label{Iw0}
I_{\omega_0}
=
- \frac{1}{\pi T} 
(1 + e^{\omega_0/T}) 
\int_0^1 dx \, \frac{1-x^n}{1-x} \, x^{-i \varphi_0 - 1/2}.
\end{eqnarray}

\subsubsection{Method I}

The  integral in (\ref{Iw0}) can be evaluated using \cite{GR2}, giving
\begin{eqnarray}
I_{\omega_0}
=
- \frac{1 + e^{\omega_0/T}}{\pi T}
\left[
	\psi \left( n - i \varphi_0 + \frac{1}{2} \right)
\right.
\\
\nonumber
\left.
	-
	\psi \left( - i \varphi_0 + \frac{1}{2} \right)
\right],
\end{eqnarray}
where
$\psi(w)$
is the digamma function.

Ultimately, we need the real-frequency propagator. To obtain it, let us
perform the substitution
$i \omega_n \rightarrow \omega + i0$,
or, equivalently,
$n \rightarrow \omega/i 2 \pi T + 0$.
In the real-frequency domain the following holds
\begin{eqnarray}
I_{\omega_0}
=
- \frac{1 + e^{\omega_0/T}}{\pi T}
\left[
	\psi \left( 
			\frac{\omega}{i 2\pi T}
			+			 
			\frac{\omega_0}{i 2\pi T}
			+
			\frac{1}{2}
		\right)
\right.
\\
\nonumber
\left.
	-
	\psi \left( 
			\frac{\omega_0}{i 2\pi T}
			+
			\frac{1}{2}
		\right)
\right].
\end{eqnarray}
Imaginary part of 
$I_{\omega_0}$
can be expressed in terms of  elementary functions, since
\begin{eqnarray}
{\rm Im\,} \psi \left(iy + \frac{1}{2} \right)
=
\frac{\pi}{2} \tanh (\pi y).
\end{eqnarray} 
The latter equality can be derived with the help of the reflection formula
for the digamma function
$\psi (1-x) - \psi (x) = \pi \cot (\pi x)$,
in which one has to substitute
$x = 1/2 - iy$.
Consequently,
\begin{eqnarray}
\label{eq::integ}
{\rm Im\,} I_{\pm \omega_0}
=
\frac{1 + e^{\pm\omega_0/T}}{2 T}
\left[
	\tanh\left(\frac{\omega \pm \omega_0}{2T} \right)
\right.
\\
\nonumber 
\left.
	\mp
	\tanh\left(\frac{\omega_0}{2T} \right)
\right].
\end{eqnarray}

\subsubsection{Method II}

\begin{figure}[t]
\includegraphics[width=0.6\columnwidth]{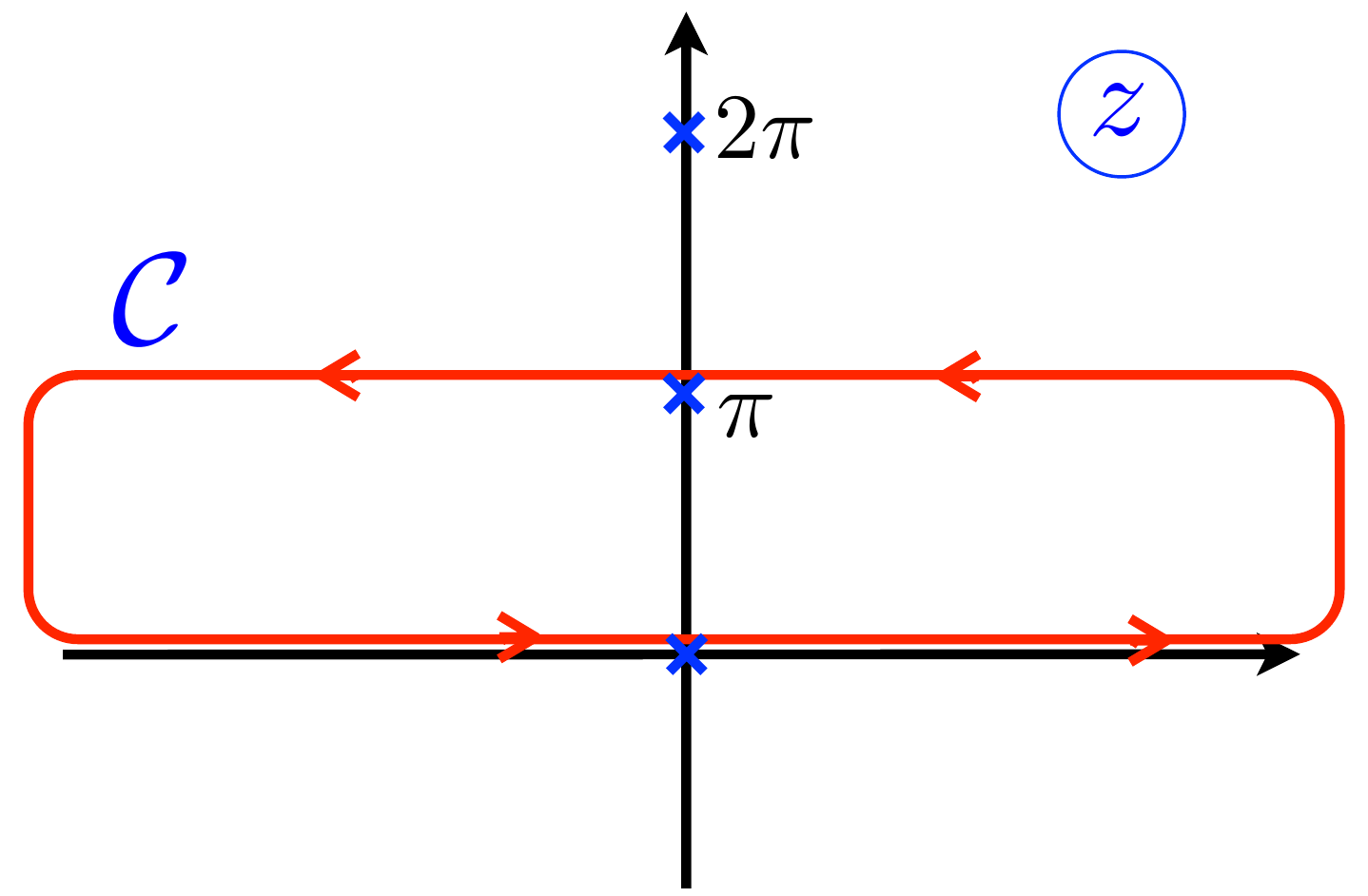}
\caption{(Color online)\ \  
Contour ${\cal C}$ for $I_{\omega_0}$ in (\ref{Iw0a}).
}
\label{Fig:contour1}
\vskip -0.3cm
\end{figure}

A substitution $x=e^{-2z}$ transforms $I_{\omega_0}$ in (\ref{Iw0}) to
\begin{eqnarray}
I_{\omega_0}
=
\frac{(1 + e^{\omega_0/T})}{\pi T} 
\int_0^\infty dz \, \frac{e^{-2nz}-1}{\sinh z} \, e^{2i\varphi_0 z}\,.
\end{eqnarray}
Since we are, ultimately, interested in the imaginary part of $\Sigma_{k}(\omega)$, it is the imaginary part 
of $I_{\omega_0}$ that we are concerned about. This is also a point at which an analytical continuation
can be made via $i \omega_n \rightarrow \omega + i0$, 
or, equivalently,
$n \rightarrow -i \omega/2 \pi T$. Introducing the variable $\varphi=\omega/2 \pi T$ and extending the 
limit to $-\infty$ we get
\begin{eqnarray}
\label{Iw0a}
&&{\rm Im\,} I_{\omega_0}
=
\frac{(1 + e^{\omega_0/T})}{\pi T} 
\\
\nonumber
&&\phantom{{\rm Im\,} I_{\omega_0}=\quad\quad}
\times\int_{-\infty}^\infty dz \, \frac{\sin (\varphi z)\cos ((2\varphi_0+\varphi)z) }{\sinh z} \,.
\end{eqnarray}
The resulting integral can be found in \cite{GR3} or evaluated by using integration in a complex plane with the help of a
standard trick utilizing the contour in Fig.~\ref{Fig:contour1}.
Both yield identical results, coinciding with
Eq.~(\ref{eq::integ}).
The quantity
${\rm Im\,} I_{-\omega_0}$
can be obtained from the latter result by simple substitution
$\omega_0\rightarrow -\omega_0$.

\subsubsection{Scattering rate}

The imaginary part of the self-energy in (\ref{Sigma_om}) is proportional
to a combination of ${\rm Im\,} I_{\pm\omega_0}$ with the phonon
distribution functions that correspond to the phonon absorption or emission
processes
\begin{eqnarray}
F(\omega,\omega_0)
=
n_0 {\rm Im\,} I_{\omega_0}
+
(n_0 + 1){\rm Im\,} I_{-\omega_0} \, ,
\end{eqnarray}
where
$n_0$
is defined in
Eq.~(\ref{eq::n0_def}).
The expression for
$F(\omega, \omega_0)$
can be rewritten as a combination of the phonon and effectively fermionic
distribution functions
\begin{eqnarray}
F(\omega,\omega_0)=\frac{1}{T}\,(1+2n_0)(1-f_+ -f_-) \, ,
\\
\mbox{with} \quad f_\pm=\frac{1}{e^{\omega\pm\omega_0}+1}\, .
\end{eqnarray} 
All of that finally yields the imaginary part of the (retarded) self-energy
\begin{eqnarray}
\label{SigmaIm}
{\rm Im\,}\Sigma_{k}^{\rm ret}(\omega)
=
-g_{\rm sp}^2\, \frac{2J}{a |k|}\,(1+2n_0)(1-f_+ -f_-)
\, ,
\end{eqnarray} 
where the dimensionless spin-phonon coupling constant
$g_{\rm sp}\!=\!\lambda/(aJ\sqrt{2 m \omega_{0}})$
is introduced.

We would like to point out that the only approximation that has been made
in the derivation of our answer in
(\ref{SigmaIm}) from the Matsubara self-energy in (\ref{Sigma_s}) 
is the omission of the ${\cal O}(T/J)$ terms in Eq.~(\ref{eq::spin_corr}).

At $\omega<T$ one can approximate the imaginary part of the self-energy in
(\ref{SigmaIm})
as
\begin{eqnarray}
\label{SigmaImapp}
{\rm Im\,}\Sigma_{k}^{\rm ret}(\omega)
\approx
- g_{\rm sp}^2\, \frac{2J\omega}{a |k|T}
\cdot\frac{1}{\sinh\left(\omega_0/T\right)}
\, .
\end{eqnarray} 
We   have verified the validity of   approximation in  (\ref{SigmaImapp}) by
a numerical calculation of the relaxation rate using (\ref{SigmaIm}) directly. The results differ by a slowly varying 
$T$-dependent function with the values near unity.

Finally, the scattering rate of spin-bosons on phonons
is obtained by taking ${\rm Im\,}\Sigma_{k}^{\rm ret}(\omega)$
on the mass surface, $\omega = v |k|$, 
\begin{eqnarray}
\label{Gamma_s}
\Gamma^{\rm 1ph}_k
= - {\rm Im\,} \Sigma_{k}^{\rm ret} (v |k|) 
=g_{\rm sp}^2\, \frac{\pi J^2}{T}\cdot\frac{1}{\sinh\left(\omega_0/T\right)}
\, ,
\end{eqnarray} 
where 
$v = \pi J a/2$
was used as above.
This single-phonon scattering rate
vanishes exponentially as a function of temperature for $T< \omega_0$
and saturates at a constant value for $T > \omega_0$.

\section{Two-phonon scattering}

Here we extend the formalism developed above to the interaction 
of spin-boson with two phonons. In the limit of purely
Einstein phonons this procedure is straightforward. We model the two-phonon
coupling  as 
\begin{eqnarray}
H_{\rm 2ph} =
\frac{\lambda^{\rm 2ph}}{a^2}
\!\!\!
\sum_{\alpha = x,y,z} 
\int \!\! dx {\bf U}_\alpha^2
	\left(
		\psi^\dag_{\rm L} \psi^{\vphantom{\dagger}}_{\rm R}
		e^{- i \pi x/a}
		+
		{\rm h.c.}
	\right),
\quad
\\
\mbox{where} \quad \lambda^{\rm 2ph}
=
a^2
\frac{\partial^2 J}{\partial r_x^2}.
\qquad
\end{eqnarray}
Needless to say, this Hamiltonian represents a greatly simplified version of the general situation 
as the coupling constant
$\lambda^{\rm 2ph}$ is  independent of both the branch index $\ell$ and polarization
$\boldsymbol{\xi}_\ell$ of a phonon.  
The square of the phonon field, which enters this Hamiltonian, equals to
\begin{eqnarray}
{\bf U}_\alpha^2 (x)
=
\frac{a}{L}
\sum_{q, \ell, q', \ell'}
	\frac{e^{i(q+q') x}}{2 m \omega_0}
	\boldsymbol{\xi}^\alpha_{q\ell}
	\boldsymbol{\xi}^\alpha_{q'\ell'}
	(a^\dag_{q\ell} + a^{\vphantom{\dagger}}_{-q\ell})
\\
\nonumber
\times
	(a^\dag_{q'\ell'} + a^{\vphantom{\dagger}}_{-q'\ell'}).
\end{eqnarray}
According to the Wick's theorem, the corresponding propagator can be
expressed as a product of two single-phonon propagators
\begin{eqnarray}
\label{2ph_propag}
\left\langle
	{\bf U}_\alpha^2 (0,0) 
	{\bf U}_\beta^2 (x,\tau) 
\right\rangle
=
2
\left\langle
	{\bf U}_\alpha (0,0) 
	{\bf U}_\alpha (x,\tau)
\right\rangle^2
\delta_{\alpha\beta}.
\end{eqnarray} 
The propagator itself is given by 
Eq.~(\ref{ph_propag}).
The singularity of the form
$[a \delta(x)]^2$
in 
(\ref{2ph_propag}) 
can be resolved through the usual prescription:
$[a \delta(x)]^2
=
a \delta(x)$.
Consequently
\begin{eqnarray}
&&\left\langle
	{\bf U}_\alpha^2 (0,0) 
	{\bf U}_\beta^2 (x,\tau) 
\right\rangle
\\
\nonumber
&&\phantom{\langle {\bf U}}
=
\frac{a \delta (x)}{2 m^2 \omega_{0}^2}
	\left[
		n_0 e^{\omega_{0} \tau}
		+
		(n_0 + 1) e^{-\omega_{0} \tau}
	\right]^2
\delta_{\alpha\beta}.
\end{eqnarray}
We note that for a multi-atomic unit cell the
latter expression has to be multiplied by 
$N_a^2$. In other words, the multi-atomic unit cell enhances the
contribution of the multi-phonon scattering.

Applying the same technical approach used in the single-phonon calculation
above to the two-phonon case is quite straightforward. The self-energy is 
\begin{eqnarray}
&&\Sigma_{k}^{\rm 2ph}(\omega)
=-
\frac{(\lambda^{\rm 2ph})^2}{\pi a^6 |k|}
\frac{3a}{2 m^2 \omega_{0}^2}
\frac{\pi T}{J }
\\
\nonumber 
&&\phantom{\Sigma_{k}^{\rm 2ph}}
\times
\left[
	n_0^2 I_{2 \omega_0}
	+
	2 n_0(n_0 + 1) I_0
	+
	(n_0 + 1)^2 I_{-2 \omega_0}
\right],
\end{eqnarray} 
where 
$I_{\pm 2 \omega_0, 0}$
are defined according to 
Eq.~(\ref{eq::def_integral}). Then, one can obtain the scattering rate  from this expression
\begin{eqnarray}
\Gamma^{\rm 2ph}
\propto
\frac{(\lambda^{\rm 2ph})^2 v}{J m^2 a^5 \omega_0^2}
\frac{ \cosh(\omega_0/2T)}{T \sinh^2 (\omega_0/T)}.
\end{eqnarray}
Using  $v \!=\! \pi a J/2$  and introducing dimensionless spin-two-phonon coupling constant,
$g_{\rm sp,2}^2\!=\!(\lambda^{\rm 2ph}/J)^2/(2m a^2 \omega_0)^2$,
we simplify the last expression to
\begin{eqnarray}
\Gamma^{\rm 2ph}
\propto
g_{\rm sp,2}^2\, \frac{J^2}{T}\cdot
\frac{ \cosh(\omega_0/2T)}{\sinh^2 (\omega_0/T)}\, .
\end{eqnarray}
One can relate the two-phonon to the one-phonon  coupling constant, 
$g_{\rm sp,2}^2\!\propto\!C_2g_{\rm sp}^4$,
where
$C_2$
is a large combinatorial factor due to multiple phonon modes that can be
involved in the scattering. At
$T\!<\!\omega_0$,
the two-phonon scattering rate yields the same exponential behavior as
(\ref{Gamma_s}).
However, at
$T\!>\!\omega_0$,
it carries  an extra power of
$T/\omega_0$ 
\begin{equation}
\Gamma^{\rm 2ph}
\propto g_{\rm sp, 2}^2 \frac{J^2T}{\omega^2_0}\, ,
\end{equation}
thus giving a natural  expansion in $T/\omega_{0}$ in the multi-phonon scattering of spin-bosons.

\section{Comment on prior work}

The spin-phonon scattering considered in Ref.~\onlinecite{CRs}, while similar to (\ref{H_U}), 
included coupling only to acoustic phonons, which, in turn, necessarily lead only to a 
small-momentum scattering of spin excitations, with the corresponding scattering rate carrying 
a smallness of a high power of $T/J\ll 1$ compared with the large-momentum
scattering of the present work. As a result, the mechanism of \cite{CRs}
required large spin-phonon coupling constant, $g_{\rm ac}\sim 10$,
to explain experimental data. If the realistic coupling is used, the
scattering processes studied
in~\cite{CRs}
turn out to be much weaker than those of the present work for most
temperatures. 

However, at low temperatures, $T\ll \Theta_D$, the mechanism discussed in \cite{CRs} can still 
be relevant as the exponentially small  population of  optical phonons in (\ref{Gamma_s}) competes with 
the power-law smallness of the small-momentum transfer mechanism of \cite{CRs}. 
In the notations of the present work, the scattering rate due to  
acoustic phonons at low-$T$ can be written as \cite{CRs} 
\begin{eqnarray}
\frac{1}{\tau_{\rm ac}}\propto g_{\rm ac}^2 \, \frac{T^5}{J^4\Theta_D} ,
\end{eqnarray}
where $g_{\rm ac}$ is the \emph{dimensionless} coupling constant as defined in the present work.
Making assumption on coupling constants being the same and $\Theta_D=\omega_0$, one
can compare this result with (\ref{Gamma_s}) to find the temperature $T^*$ at which contributions of the 
two mechanisms to the scattering are equal. 
For the large-$J$ systems, taking $\Theta_D/J\approx 0.1$ leads to $T^*\approx 0.1\Theta_D$. 
This estimate puts the regime at which the mechanism of \cite{CRs} will be dominating 
to $T<30$K, i.e. deep inside the regime which is
controlled by impurity scattering even in the very clean materials discussed here. 
Thus, although, hypothetically, there is a regime where
the mechanism of \cite{CRs} will be dominating, it is mostly unimportant in the case of cuprates.



\begin{thebibliography}{99}
\bibitem{footnote} 
Within our theory, the $T$-dependence of the mean-free 
path for the $XY$ version of the model in Eq.~(\ref{H}) does not contain the $1/T$ prefactor in $\ell_{\rm sp}$, 
the feature that can be potentially tested experimentally or numerically.
\end{thebibliography}
\end{document}